# EFFICIENCY VERSUS INSTABILITY IN PLASMA ACCELERATORS

Valeri Lebedev[♦,1], Alexey Burov[1], and Sergei Nagaitsev[1,2]
[1]Fermi National Accelerator Laboratory, P.O. Box 500, Batavia, IL 60510
[2] Department of Physics, The University of Chicago, Chicago, IL 60637, USA

*Abstract*

Plasma wake-field acceleration is one of the main technologies being developed for future high-energy colliders. Potentially, it can create a cost-effective path to the highest possible energies for $e^+e^-$ or $\gamma$-$\gamma$ colliders and produce a profound effect on the developments for high-energy physics. Acceleration in a blowout regime, where all plasma electrons are swept away from the axis, is presently considered to be the primary choice for beam acceleration. In this paper, we derive a universal *efficiency-instability relation*, between the power efficiency and the key instability parameter of the trailing bunch for beam acceleration in the blowout regime. We also show that the suppression of instability in the trailing bunch can be achieved through BNS damping by the introduction of a beam energy variation along the bunch. Unfortunately, in the high efficiency regime, the required energy variation is quite high, and is not presently compatible with collider-quality beams. We would like to stress that the development of the instability imposes a fundamental limitation on the acceleration efficiency, and it is unclear how it could be overcome for high-luminosity linear colliders. With minor modifications, the considered limitation on the power efficiency is applicable to other types of acceleration.

## Introduction

In the recent years, the subject of plasma acceleration is of great impact and interest for the science community, as demonstrated by many publications in leading science journals [1-10]. Two basic concepts for a linear collider based on plasma wake-field acceleration (PWFA) were proposed and studied [11, 12]. In this paper we focus on several fundamental limitations on the collider beam properties, and will be mainly referring to the case, when plasma is excited by a short electron bunch, known as a drive bunch. However, most of considerations are also applicable to the plasma excitation with a short laser pulse.

Presently, the acceleration of a collider-quality electron or positron bunch in a quasi-linear plasma regime does not look feasible due to the trailing bunch interaction with plasma electrons and ions [13]. For electrons, the acceleration in a blowout (bubble) regime looks like the only alternative. For positrons, the acceleration of a collider-quality bunch does not look feasible even in a bubble regime, where (1) an absence of plasma electrons on beam axis results in strong defocusing, or (2) their presence results in a strong and detrimental interaction with plasma electrons, or (3) a complete absence of plasma (i.e. a hollow channel) near axis results in a beam break-up instability (BBU), because any external focusing is too weak to prevent it. Therefore, in this paper we focus on the limitations of the electron bunch acceleration for the "strong" bubble regime, which, we believe, is the only viable option for PWFA collider schemes. Contrary to conventional rf cavities which have very large quality factors, plasma oscillations in a bubble regime have a quality factor of about 1. In this case only one bunch can be accelerated, and the efficiency of the acceleration is determined by a fraction of energy transferred from the bubble to this bunch. Therefore, in all concepts, the trailing bunch is placed behind the drive bunch, in the same plasma bubble and is designed to absorb the maximum possible fraction of the bubble energy. In this paper, we present the *efficiency-instability* relation, which sets a limit on such an energy transfer. This limit is determined by the beam break-up (BBU) instability. We would like to stress that until ways to overcome this limit are found, plasma-based collider schemes remain impractical from the perspective of acceleration efficiency. The BBU (also known as the hose) instability in PWFA concepts has been considered previously only for drive bunches [14, 15]. Although these considerations are important, the quality of the drive

[♦] E-mail: val@fnal.gov

bunch (*i.e.* its emittance and energy spread) affects the collider luminosity only in an indirect way. In contrast, our paper is focused only on the quality of the trailing bunch. In our considerations, we assume that a bunch with an optimal longitudinal density distribution can be created. This is not necessarily true in a real accelerator. Therefore, our criterion, presented below, should be considered as the best possible outcome, not necessarily achievable in practice.

## Driving a Plasma Wave

The strong bubble regime is such that the plasma bubble size, $R_b$, is much larger than the plasma shielding radius, $R_b \gg k_p^{-1}$, where $k_p = \omega_p / c = (4\pi n_0 r_e)^{1/2}$, $\omega_p$ is the plasma frequency, $n_0$ is the plasma density, $c$ is the speed of light, and $r_e$ is the classical electron radius. In this case, the dependence of a radial bubble size, $r_b$, on the longitudinal coordinate related to the bunch $\xi = ct - z$ is well approximated by the Lu equation [2]:

$$r_b \frac{d^2 r_b}{d\xi^2} + 2\left(\frac{dr_b}{d\xi}\right)^2 + 1 = \frac{2}{\pi n_0 r_b^2} \frac{dN_d}{d\xi}, \quad (1)$$

where $dN_d/d\xi$ is the linear particle density of a bunch. The longitudinal electric field on the bubble's axis is [2]

$$E_\parallel = -2\pi n_0 e r_b \frac{dr_b}{d\xi} \quad (2)$$

The Lu equation was analytically solved in Ref. [5] for a constant deceleration force along the drive bunch; some of those formulas are reproduced here for the reader's convenience.

In the case when bunch particles are absent inside the bubble, $dN_d/d\xi=0$, an integration of Eq. (1) yields the bubble shape and the electric field inside the bubble:

$$\frac{dr_b}{d\xi} = \pm\sqrt{\frac{1}{2}\left(\frac{R_b^4}{r_b^4} - 1\right)}; \; E_\parallel = \pm \pi e n_0 r_b \sqrt{2\left(\frac{R_b^4}{r_b^4} - 1\right)}, \quad (3)$$

where $R_b$ is the maximum bubble radius. Consequently, the half-bubble length is equal to:

$$\xi_b = \sqrt{2}\int_0^{R_b}\left(\frac{R_b^4}{r^4} - 1\right)^{-1/2} dr \approx 0.847 R_b. \quad (4)$$

The shape of the bubble can be approximated by

$$r_b \approx R_b \sqrt[3]{1 - \left(\frac{\xi}{\xi_b}\right)^2} \quad (5)$$

As pointed out in Ref. [5], the solution with a constant decelerating field along the drive bunch is especially interesting since the energy spread of the bunch would be minimized. Assuming that all particles in the bunch are decelerated with the same rate, $E_d$, the integration of Eq. (2) relates the bubble radius at the location $\xi$ and the electric field:

$$E_d \xi = \pi n_0 e r_b^2, \quad (6)$$

where $\xi$ is measured from the head of the bunch and we accounted that $r_b=0$ at the driving bunch head. Substituting $r_b$ into Eq. (1), one obtains the corresponding longitudinal distribution:

$$\frac{dN_d}{d\xi} = \frac{E_d}{8\pi n_0 e}\left(E_d / e + 4\pi n_0 \xi\right), \quad (7)$$

leading to the total number of particles in the bunch

$$N_d = \frac{E_d L_d}{8\pi n_0 e}\left(E_d / e + 2\pi n_0 L_d\right), \quad (8)$$

where $L_d$ is the full drive bunch length. According to Eq. (6), $E_d L_d = \pi e n_0 r_d^2$, where $r_d$ is the bubble radius at the drive bunch end. Due to continuity of the deceleration field at the end of the bunch,

$$E_d = \pi e n_0 r_d \sqrt{2\left(\frac{R_b^4}{r_d^4} - 1\right)}. \quad (9)$$

Using these equations, the bunch power losses are easily obtained:

$$P = eN_d E_d c = \frac{\pi^2}{4} e^2 n_0^2 c R_b^4. \quad (10)$$

Thus, the maximum bubble radius uniquely determines the power transferred by the drive bunch to plasma.

The constancy of the decelerating force along the bunch contradicts the general requirement for the decelerating force to be zero for the very head of the relativistic drive bunch. This contradiction originates from the poor description of the plasma reaction for small $r_b$ by the Lu equation; the equation works well only for $L_d \geq k_p^{-1}$. When the bunch is sufficiently long, the inaccuracy of the Lu equation at the very head of the bunch is insignificant for the bubble formation.

For a given number of particles, the decelerating electric field can be obtained from Eq. (8). It is also straightforward to obtain the maximum bubble radius:

$$R_b = \frac{L_d}{\sqrt[4]{2}}\sqrt[4]{\frac{8N_d}{\pi n_0 L_d^3}\left(\sqrt{\frac{8N_d}{\pi n_0 L_d^3} + 1} - 1\right)}, \quad (11)$$

from the above equations. Eq. (11) can be approximated by the following:

$$R_b \approx \left(\frac{2^7 N_d^3}{\pi^3 L_d n_0^3}\right)^{1/8}, \quad \frac{N_d}{n_0 L_d^3} \gg 1. \quad (12)$$

One can notice that the bubble radius diverges at small bunch lengths. Although the divergence is quite slow, it still shows that for a delta-function drive bunch the Lu equation is mathematically incorrect.

## Acceleration

Now let us consider the acceleration of a trailing bunch. Similarly to the drive bunch, the particle density of the trailing bunch can be chosen such that all particles are accelerated at the same rate. If so, the trailing linear density is trapezoidal, like the one described by Eq.(7), linearly decreasing toward the bunch tail [5]. Expressing coordinates of the bunch head and tail through the bubble radii at their locations, the total number of particles in the trailing bunch is:

$$N_t = \frac{E_t}{8e}\left(r_{t2}^2 - r_{t1}^2 + 2\left(\frac{\pi n_0 e}{E_t}\right)^2 (r_{t2}^4 - r_{t1}^4)\right). \quad (13)$$

Here $E_t$ is the accelerating field; $r_{t1}$ and $r_{t2}$ are the bubble radii in the locations of bunch tail and head, respectively; also we took into account that the trailing bunch length $L_t = \pi e n_0 (r_{t2}^2 - r_{t1}^2)/E_t$. Then, the power transferred to the trailing bunch is:

$$P_t = ecN_tE_t = \frac{\pi^2 e^2 n_0^2 c}{4}(r_{t2}^2 - r_{t1}^2)\left(\frac{R_b^4}{r_{t2}^2} + r_{t1}^2\right). \quad (14)$$

Consequently, the efficiency of power transfer from a drive to a trailing bunch is:

$$\eta_P = \frac{P_t}{P} = \frac{r_{t2}^2 - r_{t1}^2}{R_b^2}\left(\frac{R_b^2}{r_{t2}^2} + \frac{r_{t1}^2}{R_b^2}\right). \quad (15)$$

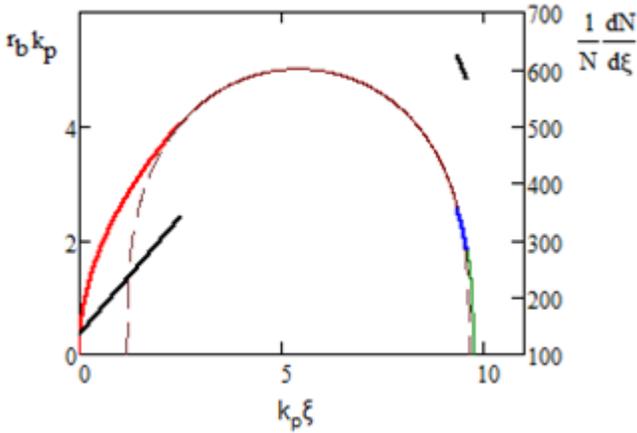

**Figure 1:** The bubble shape and particle distributions (black lines) for drive and trailing bunches, with $n_0=10^{17}$ cm$^{-3}$. The red and blue lines mark the parts of the bubble, occupied by the drive and trailing bunches, respectively. The dashed brown line shows the analytical continuation of the bubble shape in the absence of bunch particles inside the bubble. The accelerating and decelerating fields are constant: $E_d$=50 GV/m, $E_t$=100 GV/m.

Figure 1 shows an example, illustrating the bubble shape and the particle distributions of the drive and trailing bunches for the power transfer efficiency of 50% and the transformer ratio $E_t / E_d$ of 2. For $n_0=10^{17}$ cm$^{-3}$ the drive bunch parameters are chosen to be $R_bk_p$=5, $L_dk_p$=2.5 yielding the decelerating field of $E_d$ = 50 GV/m and $N_d$=3.55·10$^{10}$. The trailing bunch parameters are: $r_{t2}$=0.518$R_b$, $r_{t1}$=0.373$R_b$, $E_t$= 100 GV/m, $N_t$=8.86·10$^9$.

## Instability

The *Beam Break-up* (BBU) instability is characterized by the ratio of the wake deflection force to the focusing force. The latter is given by

$$F_r = -2\pi n_0 e^2 r. \quad (16)$$

where $r$ is the particle offset from the bubble axis.

In the strong bubble regime all plasma currents are localized in a thin layer near the bubble boundary. In this case the transverse and longitudinal wakes are related to each other by a universal expression [16, 17], which in further considerations we will call the short-range wake theorem (see also Ref. [17] and multiple references therein):

$$W_\perp(\tilde{\xi}) \approx \frac{2}{\tilde{r}_b^2}\int_0^\xi W_L(s)ds, \quad (17)$$

where

$$\tilde{r}_b = r_b + k_p^{-1} \quad (18)$$

is the effective bubble radius at the driving particle location, $\tilde{\xi} = \xi - \xi_2$ is the distance between leading and trailing particles, and $\xi_2$ and $\xi$ are the positions of the leading and the trailing particles in the trailing bunch. The $k_p^{-1}$ correction term in Eq. (18) accounts for a finite penetration depth of the beam induced currents into a plasma. In the case of a hollow plasma channel [18], which is solved analytically, such a correction makes Eq. (17) to be a good approximation to an exact solution for $r_b k_p \geq 3$. Note that plasma ions inside the bubble do not make a considerable contribution to the wakes, because of their low mobility, compared to the mobility of electrons.

The short-range wake theorem is usually applied to structures with a solid aperture, where its application has been well justified by many authors. It was shown to be correct for the resistive wall [19] with thin skin depths, for dielectric-covered pipes, for pipes with small corrugations [17]. If the wall conductivity is sufficiently high, the result does not depend on the conductivity. This indicates that the use of the short-range wake theorem is justified for a plasma bubble (or a plasma channel), if all plasma currents are concentrated in a layer with the thickness much smaller than the channel radius. This condition is identical to the condition of the strong bubble regime, $R_b \gg 1/k_p$, considered here.

It is straightforward to obtain the longitudinal wake immediately behind a test particle. Placing a point-like charge $q\delta(\xi-\xi_2)$ on the axis, integrating Eq.(1) in the close vicinity of $\xi_2$ and, then, substituting the result into Eq. (2) one obtains:

$$W_L = \frac{4}{r_b^2}, \qquad (19)$$

where $r_b$ is the bubble radius at the particle location. As one can see, unlike Eq. (18), the bubble radius in Eq. (19) does not have an addition of $k_p^{-1}$. This is determined by the applicability condition of the Lu equation: $R_b k_p \gg 1$. The numerical integration of Eqs. (1) and (2) yields an approximation for the wake function, which describes the wake sufficiently accurately almost to the end of the bubble:

$$W_L(\xi,\xi_2) \approx \frac{4}{r_b(\xi)^2}\theta(\xi-\xi_2), \qquad (20)$$

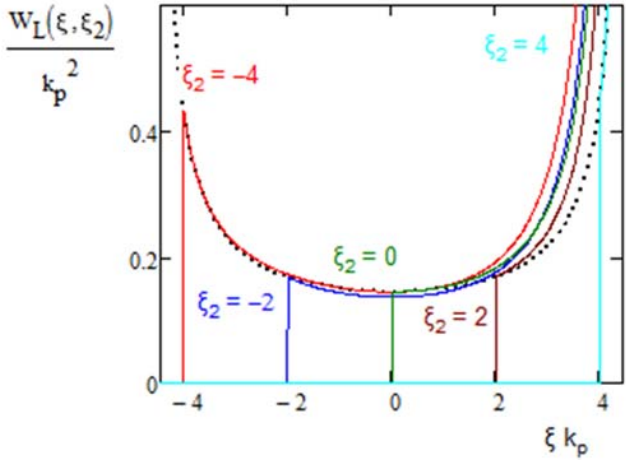

**Figure 2:** Dependence of the longitudinal wake on the longitudinal coordinate $\xi$ for different locations of the leading particle $\xi_2 = -4, -2, 0, 2, 4$; $R_b k_p = 5.25$, $\xi_b k_p = 4.448$. The dashed line shows the function $4/(r_b(\xi) k_p)^2$ which is directly related to the wake wake-function of Eq. (20).

where $\theta(x)$ is the Heaviside step function. Figure 2 presents a comparison of the wake-functions of Eq. (20) and obtained by numerical integration. As one can see each wake function starts from the value prescribed by Eq. (20) and slowly diverges from it near the bubble end. Note that this dependence is quite different from the case of soft plasma excitation where the wake oscillates at a frequency close to the plasma frequency.

All mentioned above applications of the short-range wake theorem are related to channels of constant radius. It is not entirely obvious how this theorem has to be applied to a plasma bubble, which radius is changing with coordinate $\xi$. For this paper, we assume that the wake is equal to:

$$W_\perp(\xi,\xi_2) \approx \frac{8\tilde{\xi}}{r_b(\xi)r_b^3(\xi_2)}\theta(\tilde{\xi}), \quad \tilde{\xi} = \xi-\xi_2. \quad (21)$$

This result follows from Eqs. (17) and (20) in the leading order of expansion over the bunch length. The actual value of the transverse wake is expected to be somewhat larger because, like the longitudinal wake, the transverse wake should be mostly dependent on the bubble radius at the trailing particle position. Thus, our choice of the transverse wake potential yields an optimistic value for the instability threshold. The exact value of the transverse wake should be a subject of another study.

Assuming that initially all particles of the trailing bunch are off-axis at the same radius, $r$, one obtains the wake force acting on particles at the bunch tail:

$$F_t \equiv F(\xi_1) = e^2 r \int_{\xi_1-L_t}^{\xi_1} \frac{dN_t}{d\xi} W_\perp(\xi_1,\xi)d\xi, \qquad (22)$$

where $L_t$ is the length of the trailing bunch, and $\xi_1$ is the longitudinal coordinate of the bunch tail. The ratio of the wake-deflecting force to the focusing force of Eq. (16), or the normalized wake defocusing, is:

$$\eta_t = -\frac{F_t}{F_r} = \frac{r_{t2}}{r_{t1}}\int_0^{L_t} d\xi \frac{L_t-\xi}{r_b^3(\xi)} \times$$
$$\left[r_{t2}\left(\frac{R_b^4}{r_{t2}^4}-1\right)-2\left(\xi\sqrt{2\left(\frac{R_b^4}{r_{t2}^4}-1\right)}-r_{t2}\right)\right]. \qquad (23)$$

Here we changed the origin of longitudinal coordinate so that $\xi = 0$ at the bunch head and it grows to the bunch tail,

$$r_b(\xi) = \sqrt{r_{t2}^2 - r_{t2}\xi\sqrt{2\left(\frac{R_b^4}{r_{t2}^4}-1\right)}},$$

$$L_t = \frac{r_{t2}^2 - r_{t1}^2}{r_{t2}}\left(2\left(\frac{R_b^4}{r_{t2}^4}-1\right)\right)^{-1/2}, \qquad (24)$$

and we accounted that the longitudinal density in the accelerated bunch is:

$$\frac{dN}{d\xi} = \frac{E_t}{8\pi e n_0}\left(\frac{E_t}{e} - 4\pi n_0\left(\xi - \frac{\pi e n_0 r_{t2}^2}{E_t}\right)\right). \quad (25)$$

Eqs. (24) follow from Eqs. (1) - (3) and Eq. (25) is obtained similar to Eq. (7).

Plotting $\eta_t$ as function of $\eta_P$ for various values of $r_{t1}$ and $r_{t2}$ one obtains that in the area of interest, $r_{t2}/R_b \leq 0.7$, where the acceleration is reasonably fast, the normalized wake defocusing, $\eta_t$, can be related to the power transfer efficiency:

$$\eta_t \approx \frac{\eta_P^2}{4(1-\eta_P)}, \quad \frac{r_{t2}}{R_b} \leq 0.7. \qquad (26)$$

Note that this formula does not include any details of beams and plasma, being amazingly universal. Due to its importance, simplicity and universality, we propose to name it the *efficiency-instability relation*. The term in the denominator is determined by the specific dependence of the transverse wake on $\xi$, while $\eta_P^2/4$ is universal and is applicable to any structure. In Ref. [13] we considered the opposite limiting case of the transverse wake behavior when the wake is determined by the bubble radius at the location of the trailing particle only: $W_\perp(\xi,\xi_2) \approx 8\xi\theta(\xi)/r_b^4(\xi)$. In this case, the instability-efficiency relation of Eq. (26) has the following form:

$$\eta_t \approx \eta_P^2 / \left(4(1-\eta_P)^2\right)$$

In the following analysis we will neglect the line density variation along the trailing bunch, assuming this density is not trapezoidal, but rectangular. This approximation is reasonable when the trailing bunch is not too close to the end of the bubble. In that case, the line density variation is small enough, like in Figure 1. It also implies that in Eq. (21) $r_b(\xi) \approx r_b(\xi_2)$, and the power efficiency is sufficiently small so that Eq. (26) is transformed into $\eta_t \approx \eta_P^2/4$.

Strong focusing in the bubble results in a large number of betatron oscillations during the beam acceleration. The total betatron phase advance can be estimated as:

$$\mu = \sqrt{2}\left(\sqrt{\gamma_f} - \sqrt{\gamma_i}\right)E_0/E_t; \quad E_0 = 4\pi n_0 e/k_p, \quad (27)$$

where $\gamma_f$ and $\gamma_i$ are the final and initial values of the Lorentz factor. In this case, the oscillations of the bunch head resonantly drive particles in the tail, resulting in an increase of the effective transverse emittance.

To describe this head-tail motion, we will use the normalized variables:

$$X = \frac{x}{\sqrt{\beta}}\sqrt{\frac{p}{p_0}}; \quad \beta = k_p^{-1}\sqrt{2\gamma}. \quad (28)$$

With $d\mu = dz/\beta$, an equation for the transverse oscillations is:

$$\frac{d^2X}{d\mu^2} + \frac{X}{1+\Delta p/p} = \frac{2\eta_t}{(1+\Delta p/p)L_t^2}\int_0^\xi X(\xi')(\xi-\xi')d\xi'. \quad (29)$$

Here $p$ and $p_0$ are the momentum and its initial value; $\Delta p/p$ is a possible momentum deviation as a function of the intra-bunch coordinate, $\xi$, and we assume that radii of the driving and leading particles in the expression for the transverse wake of Eq. (21) are equal.

Let all trailing particles have the same initial normalized amplitude, $X = A_0$, resulting from an offset between the axes of the drive and trailing bunches. For $\eta_t \ll 1$ and $\Delta p/p = 0$, Ref. [18] presents an asymptotic solution of Eq. (29) for $\mu\eta_t \gg 1$. To obtain a solution for practical phase advances, we solved the equation numerically. The obtained results suggest an approximate parameterization for the ratios of the tail particle amplitude

$$\frac{A}{A_0} = \exp\left(\frac{(\mu\eta_t)^2}{10+1.4(\mu\eta_t)^{1.57}}\right); \quad \begin{array}{l}\mu\eta_t \leq 100 \\ \eta_t \leq 0.1\end{array}, \quad (30)$$

and the rms amplitude averaged over all particles

$$\frac{\sqrt{\overline{A^2}}}{A_0} = \exp\left(\frac{(\mu\eta_t)^2}{60+2.2(\mu\eta_t)^{1.57}}\right); \quad \begin{array}{l}\mu\eta_t \leq 100 \\ \eta_t \leq 0.1\end{array}. \quad (31)$$

to the initial amplitude. These approximations do not deviate by more than 10% from the numerical solution in the range of interest. Figure 3 presents the corresponding plots. The rms amplitude of Eq. (31) is determined as

$$\sqrt{\overline{A^2}} \equiv \sqrt{\frac{1}{N}\sum_{n=1}^{N}A_n^2},$$

where averaging is performed for all particles, and $A_n$ is the betatron amplitude of $n$-th particle.

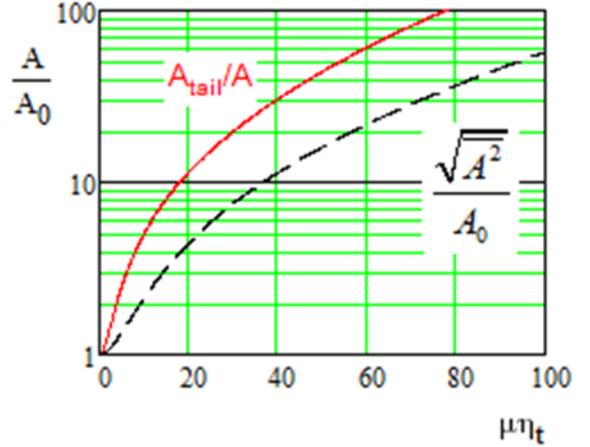

**Figure 3:** The dependencies of ratios for the tail amplitude particles (top solid line) and the rms amplitude of all particles (bottom dashed line) to the initial amplitude, $A_0$.

As our first example, we will consider a single-stage 60 cm long plasma section with $n_0 = 4\times 10^{16}$ cm$^{-3}$, initial momentum $p_i$=10 GeV/c for both the drive and the trailing bunches, and the final momentum of trailing bunch $p_f$=21 GeV/c, $N_d$=1x10$^{10}$ and $N_t$=4.3x10$^9$. These parameters are of specific interest for the future FACET-II research program [20]. We will consider two somewhat arbitrary examples. For the power transfer efficiency of $\eta_P = 50\%$, we obtain $\mu\approx 93$ rad from Eq. (27) and $\eta_t \approx 0.12$ from Eq.(26). A use of Eqs. (30) and (31) yields the amplitude growth for the tail particles $A/A_0 \approx 5.7$, and the relative rms amplitude $\sqrt{\overline{A^2}}/A_0 \approx 2.3$. A displacement of the trailing bunch closer to the bubble center reduces the accelerating rate and, consequently, the power efficiency. If we reduce the power efficiency by a factor of two to $\eta_P = 25\%$, the energy gain is also reduced by a factor of 2 (for the same plasma length and particle number), $p_f$=15.5 GeV/c and the amplitude growths become: $A/A_0 \approx 1.35$, $\sqrt{\overline{A^2}/A_0^2} \approx 1.07$. Note that the corresponding increase of the normalized emittance is

$$\delta\varepsilon_n = \frac{\delta x^2}{2\beta_i}\gamma_i\left(\frac{\overline{A^2}}{A_0^2}\right), \quad \beta_i = \frac{\sqrt{2\gamma_i}}{k_p}, \quad (32)$$

where $\delta x$ is the transverse offset of the trailing bunch relative to the driving bunch, and $\beta_i$ is the beta-function at the beginning of accelerating section.

As the second example, let us consider a 1-TeV linac with $\mu \approx 10^3$. In the case of a single kick, tolerance to the amplitude growth is more forgiving. However, it would be more realistic to expect many perturbations to the machine alignment coming from ground motion, jitter in the driving beam position or positions of laser beams in the case of laser-plasma acceleration, etc. If so, a single offset should not increase the trailing particle squared amplitude by, say, by more than an order of magnitude, yielding $\mu\eta_t < 10$ and thus $\eta_t < 0.01$. Using the efficiency-instability relation Eq. (26), we obtain a limitation on the energy transfer $\eta_P < 18\%$.

An effective way to suppress the BBU instability, the BNS damping, was suggested by Balakin, Novokhatsky and Smirnov [21]. The idea is to introduce a dependence of particle momentum on the longitudinal coordinate, $\xi$, in the bunch so that it would compensate frequency detuning due to transverse wake. To accomplish that, Eq. (29) requires:

$$\frac{1}{1+\frac{\Delta p}{p}} - \frac{2\eta_t}{\left(1+\frac{\Delta p}{p}\right)L_t^2}\int_0^\xi (\xi-\xi')d\xi' = 1. \quad (33)$$

That results in:

$$\frac{\Delta p(\xi)}{p} = -\eta_t \frac{\xi^2}{L_t^2}. \quad (34)$$

For colliders, estimates of chromatic aberrations in the final focus suggest that the total momentum spread can hardly be allowed to exceed 1% [22] (see also discussion in [13]). If so, it yields the same value of $\eta_t \leq 0.01$ and, by virtue of the efficiency-instability relation (26), it sets the same limit on the power efficiency as without BNS damping. Note also that it is unclear how the quadratic dependence of momentum deviation on $\xi$ required by Eq. (34) can be created for the entire length of the accelerator.

A remedy for mitigation of the hose instability for the drive bunch was suggested in Ref. [15]. It is based on tapering of plasma density at the input and the output of a plasma channel. As will be seen below, this technique also works for the trailing bunch. A transverse misalignment of drive and trailing bunches at the plasma entrance results in an excitation of transverse oscillations of the trailing bunch with subsequent emittance growth. These oscillations represent a seed for the hose instability. The plasma density tapering produces "an adiabatic increase" of transverse focusing, which results in a reduction of betatron amplitude excited by the misalignment in the trailing bunch. The net suppression of the betatron amplitude is determined by the shape and the length of plasma tapering. For the case when the density transition length, $L_{tr}$, is much larger than the beta-function of the transverse motion in plasma, $\beta$, the value of suppression is about $\sqrt{L_{tr}/\beta}$. This suppression can be quite significant at low beam energies, where $\beta$ is small. With energy increase, $\beta$ also increases (see Eq. (28)). This results in that the required transition length in the second half of 1 TeV accelerator is more or about 10 cm, which can be difficult to achieve in practice. Finally, we would like to point out that plasma tapering can be helpful to reduce requirements on the misalignment of the drive and trailing bunches but does not change the derived above limitations on the energy transfer efficiency, since tapering does not change the development of the instability inside the plasma itself.

### Other limitations

While plasma-based acceleration of intense high-quality electron bunch is feasible, albeit challenging, the same cannot be said about positron bunches. For the positron acceleration, the plasma electron density on the bunch axis needs to exceed that of ions for sufficiently strong focusing to counteract the transverse BBU instability. The transverse wake field is so large that the instability suppression cannot be obtained by any means other than plasma focusing. However, an introduction of plasma electrons on the axis results in a collapse of these electrons to the positron bunch center, which significantly distorts the linearity of focusing with radius [23]. Electrons within the radial size of about $r_m = \sqrt{r_e N_t L_t}$ are all pulled into the positron bunch. Even for a modest number of positrons in a bunch, $r_m$ is larger than the typical transverse beam size. The resulting high density of plasma electrons at the positron beam axis both eliminates focusing linearity and enhances multiple Coulomb scattering of positrons on plasma electrons, resulting in unacceptably large beam emittance growth. For a bunch population of $4\cdot10^9$ and a bunch length of 10 μm one obtains $r_m$=10.6 μm, while the typical transverse size is less than μm.

For an electron bunch acceleration in the bubble regime, plasma ions may also collapse in its field. This effect has been considered in Ref. [24]. In this case, $r_m$ is determined by the ion mass instead of the electron one. For above considered case with the bunch population of $8.86 \cdot 10^9$ and the bunch length of 4.2 μm one obtains $r_m$=0.2 μm for proton plasma. This size is still larger than the electron beam radius varying in the range of 0.05 – 0.15 μm. This means that a problem of the ion collapse in the field of the electron bunch is also quite severe and will be an important limitation on the collider parameters. One may also consider the phase advance of small-amplitude oscillations of plasma ions in the field of an electron bunch: $\mu_{ion} = \sqrt{2 r_p N_t L_t / (M_i \sigma_\perp^2)}$, where $M_i$ is the ratio of the ion mass to the proton mass. For the above parameters of the accelerated electron bunch and the hydrogen plasma, the proton oscillation phase advance grows with beam energy and achieves ~360° at 500 GeV. To avoid potential problems, this value needs to be reduced by at least an order of magnitude.

Although heavy-ion plasma looks as a possible means to mitigate the problem of ion collapse [9] and oscillations, its application seems to be excluded by the impact ionization of the ions. For the required bunch parameters, the electric field at its boundary exceeds ~$10^3$ GV/cm. It is more than two orders of magnitude larger than the electric field in a hydrogen atom of ~6 GV/cm. Use of ions stripped to the level sufficient to avoid impact ionization looks to be unrealistic. The use of heavy ions also increases the effects of bremsstrahlung, which are not negligible even for proton plasma.

**Discussion**

Many challenges must be overcome before a credible concept of a plasma-based $e^+$-$e^-$ or γ-γ collider can be put forward. As far as we can judge, there is still no viable path to a high luminosity collider within the present concepts. As it was already stressed, the acceleration of the required intense low-emittance positron bunch appears the most challenging.

Achieving high efficiency even in a more realistic case of a plasma-based $e^-$-$e^-$ or γ-γ accelerator represents a great challenge. It originates from a low Q-value of plasma oscillations (especially in the bubble regime), resulting in that only one bunch can be accelerated in a single pulse. The BBU instability, driven by the transverse impedance of the plasma bubble, is one of the major limitations. It limits the number of particles in the trailing bunch and, consequently, limits the efficiency of acceleration. The instability greatly amplifies the emittance growth due to errors of the relative alignment between different accelerating sections. Note that presently the required alignment accuracy of sub-μm does not look attainable even in the absence of the BBU instability. The BNS damping, which potentially could help, requires a large energy spread, which is unacceptable from the collider final focus point of view.

For present concepts, pinching of plasma ions by a bright electron beam limits the luminosity of $e^-$-$e^-$ or γ-γ collider to well below $10^{34}$ cm$^{-2}$s$^{-1}$ if a light ion plasma is used. Use of heavy ion plasma is precluded by the impact ionization by the electron bunch fields; the ions cannot be sufficiently well stripped due to the required energy efficiency. Also, multiple scattering and bremsstrahlung would be greatly amplified in that case.

In conclusion, we would like to emphasize that although there may be many applications for plasma-based accelerators, it is unclear how the limitations, described in our paper, could be overcome for high luminosity linear colliders, which would make them competitive with proposals, based on a conventional rf acceleration technology.

**Acknowledgements**

The authors are grateful to G. Stupakov for helpful and interesting discussions regarding the presented research results. This work was supported by the US DOE under contract #DE-AC02-07CH11359.